\documentclass[12pt]{article}
\usepackage{epsfig,amsfonts,amssymb}
\usepackage{hyperref}
\usepackage{cite}
\input epsf.sty
\usepackage{cite}

\def\ZZZ{{\hbox{ Z\kern-1.6mm Z}}}
\def\RRR{{\hbox{ R\kern-2.4mm R}}}
\def\CCC{{\hbox{ C\kern-2.0mm C}}}
\def\zzz{{\hbox{z\kern-1mm z}}}

\newcommand{\qeq}{{\hbox{=\kern-2.3mm ? \kern.5mm }}}
\renewcommand{\qeq}{=}

\newcommand{\FF}{{\cal F}}

\newcommand{\HH}{{\cal H}}

\newcommand{\OO}{{\cal O}}

\newcommand{\wt}{\widetilde}
\newcommand{\wh}{\widehat}

\newcommand{\be}{\begin{equation}}
\newcommand{\ee}{\end{equation}}
\newcommand{\ben}{\begin{eqnarray}\displaystyle}
\newcommand{\een}{\end{eqnarray}}

\newcommand{\refb}[1]{(\ref{#1})}
\newcommand{\p}{\partial}
\newcommand{\sectiono}[1]{\section{#1}\setcounter{equation}{0}}

\def\one{{\hbox{ 1\kern-.8mm l}}}
\def\zero{{\hbox{ 0\kern-1.5mm 0}}}

\newcommand{\bea}[1]{\begin{eqnarray}\label{#1} }
\newcommand{\eea}{\end{eqnarray}}

\newcommand{\eqref}{\refb}

\begin{document}

\begin{flushright}
DAMTP-2013-62\\
HRI/ST/1306
\end{flushright}

\vskip 12pt

\baselineskip 24pt

\begin{center}
{\Large \bf  Mass Renormalization in String Theory: Special States}

\end{center}

\vskip .6cm
\medskip

\vspace*{4.0ex}

\baselineskip=18pt

\centerline{\large \rm Roji Pius$^a$, Arnab Rudra$^b$ and Ashoke Sen$^a$}

\vspace*{4.0ex}

\centerline{\large \it ~$^a$Harish-Chandra Research Institute}
\centerline{\large \it  Chhatnag Road, Jhusi,
Allahabad 211019, India}
\centerline{\large \it ~$^b$Department of Applied Mathematics and Theoretical Physics}
\centerline{\large \it Wilberforce Road, Cambridge CB3 0WA, UK}

\vspace*{1.0ex}
\centerline{\small E-mail:  rojipius@mri.ernet.in, A.Rudra@damtp.cam.ac.uk, sen@mri.ernet.in}

\vspace*{5.0ex}

\centerline{\bf Abstract} \bigskip

String theory gives a well defined procedure for computing 
the S-matrix of BPS or a class of massless states, but similar calculation for 
general massive states is plagued
with difficulties due to mass renormalization effect. In this paper we describe a 
procedure for computing the renormalized masses and S-matrix elements in bosonic string
theory for a special class of massive states which do not mix with unphysical states under
renormalization. Even though this requires working with off-shell amplitudes which are
ambiguous, we show that the renormalized masses and S-matrix elements are free from
these ambiguities. We also argue that the masses and S-matrix elements for 
general external states can be found by examining the locations of the poles and the
residues of the  S-matrix of special states. Finally we discuss generalizations to heterotic
and superstring theories.

\vfill \eject

\baselineskip=18pt

\tableofcontents

\sectiono{Introduction} \label{sintro}

String theory does not have a non-perturbative definition at present, but it gives a well
defined procedure for computing S-matrix elements involving BPS or a class of massless
external states -- whose masses are protected from renormalization -- to 
any order in perturbation theory. Furthermore this perturbation expansion is free from
ultraviolet divergences (see {\it e.g.} \cite{1209.5461,Belopolsky,dp,Witten,1304.7798} 
for recent discussion). However the
usual procedure for computing S-matrix elements breaks down for general massive states.
This is due to the fact that for general massive states loop corrections generate (ultraviolet finite)
mass renormalizations, and hence in order to compute the physical S-matrix elements we
have to shift the external momenta to their renormalized on-shell values. On  the
other hand string perturbation theory, which is based on world-sheet conformal invariance,
requires the vertex operators representing  external states
to be dimension zero primary fields. This is 
equivalent to requiring that the external momenta satisfy the {\it tree level} on-shell
condition.

For many states this is not a problem since they appear as single particle
intermediate states in the S-matrix of massless and/or BPS 
external states and hence their renormalized
masses and S-matrix elements can be found by examining the locations and residues of
the poles of the S-matrix of massless and/or BPS states.
For this reason direct computation of the S-matrix of massive string states has not received
much attention. 
 However this does not always work,
{\it e.g.} if the massive state under consideration carries a conserved charged that
is not carried by any of the massless or BPS states, then the former cannot appear as a single particle
intermediate state in the S-matrix of the latter. For this reason it seems important to
find a more direct approach to computing the mass renormalization and S-matrix elements
of massive string states.

In this paper we undertake this task for a special class of states in bosonic string theory.
There are two related but independent problems which arise in the computation of mass
renormalization in string theory. First
we have to define the analog of the off-shell Green's function in string theory. This requires
giving up the conformal invariance of vertex operators and hence is ambiguous. Second,
under renormalization the BRST invariant physical states begin mixing with the
unphysical states of string theory and hence the definition of the physical state needs
to be modified carefully. By choosing a special class of states we avoid the second problem
-- these special class of states do not mix with unphysical states due to some global
symmetries. However we still need to deal with the first problem, \i.e.\ the ambiguity in
the definition of the 
off-shell Greens function. We show that although the off-shell Greens functions are
ambiguous, the renormalized mass and S-matrix elements computed from them are free
from these ambiguities.

The paper is organised as follows. In \S\ref{squestion} we make precise the problem associated
with mass renormalization in string theory, and also introduce the special class of states for which
we address the problem in this paper. In \S\ref{smass} we describe how to compute the renormalized
mass of these special states and also show that this renormalized mass is free from any
ambiguity. In \S\ref{smatrix} we show how to compute the S-matrix elements of these
special states, and demonstrate that they are also free from ambiguities. We end in
\S\ref{sdiscuss} with a discussion of our results,
 extensions to
heterotic and superstring theories and future generalizations.

Various other approaches to studying mass renormalization in string theory can be found in
\cite{Weinberg,Seiberg,OoguriSakai,Yamamoto,AS,Das,Rey,Sundborg,Marcus,Amano,
Lee,Berera,0812.3129,0903.3979}

\sectiono{The question} \label{squestion}

Consider a string theory amplitude with $n$-external states representing particles 
carrying momenta $k_1,\cdots k_n$ and other
discrete quantum numbers $a_1,\cdots a_n$ with
tree level masses $m_{a_1},\cdots m_{a_n}$. Then the momenta
$k_i$ satisfy the  {\it tree level} 
on-shell condition $k_i^2 = -m_{a_i}^2$, -- this is needed to ensure the
BRST invariance of the vertex operators in the world sheet theory. 
The world-sheet computation, involving correlation functions of these vertex operators
integrated over the moduli spaces of (punctured) Riemann surfaces, yields the result for what
in a quantum field theory can be called
 `truncated Green's function on
classical mass shell':\footnote{We have absored all factors of ${\bf i}\equiv \sqrt{-1}$ 
and minus signs into the
definition of $G^{(n)}$ and $R^{(n)}$.}
\ben \label{ek4}
&& R^{(n)}_{a_1\cdots a_n}(k_1,\cdots k_n) 
\equiv \lim_{k_i^2 \to -m_{a_i}^2} F^{(n)}_{a_1\cdots a_n} (k_1, \cdots k_n)
\, , \nonumber \\
&&  F^{(n)}_{a_1\cdots a_n}  (k_1, \cdots k_n)\equiv
G^{(n)}_{a_1\cdots a_n} (k_1,\cdots k_n)\prod_{i=1}^n (k_i^2 + m_{a_i}^2) \, ,
\een
where $G^{(n)}_{a_1\cdots a_n} (k_1,\cdots k_n)$ correspond to the momentum space Green's function
in the quantum field theory. 
This is similar to but not the same as the combination that appears in the expression
for the S-matrix in a quantum field theory
\be \label{ek5pre}
S^{(n)}_{a_1\cdots a_n}(k_1,\cdots k_n) =
\lim_{k_i^2 \to -m_{a_i,p}^2} G^{(n)}_{a_1\cdots a_n}
(k_1,\cdots k_n)\prod_{i=1}^n \{Z^{-1/2}(k_i, a_i) (k_i^2 + m_{a_i,p}^2)\} \, ,
\ee
where $m_{a_i,p}$ is the physical mass of the $i$-th particle, defined as the location of the pole
as a function of $-k^2$ in the untruncated two point Green's function $G^{(2)}$ and
$Z(k_i, a_i)$'s are the residues at these poles. 

For simplicity we have ignored the mixing between different states under wave-function renormalization
in writing down \refb{ek5pre},
but we shall discuss the general case now.
If we consider the set of all fields whose
tree level masses are all equal to $m$
then the two point Green's function
$G^{(2)}_{ab}(k, k')$
for all these fields is described by the matrix
\be \label{ek6}
G^{(2)}_{ab}(k, k') = (2\pi)^{D+1} \delta^{(D+1)}(k+k') \, Z^{1/2}(k)_{ac} (k^2 + M_p^2)_{cd}^{-1} 
(Z^{1/2}(-k))_{db}^T \, ,
\ee
where $M_p^2$ is the mass$^2$ matrix and $Z^{1/2}(k)$ is the wave-function renormalization matrix,
the latter being free from poles near $k^2+m^2\simeq 0$. 
The sum over $c,d$ are restricted to states which have the same tree level mass $m$ as the states
labelled by the indices $a,b$.
$D+1$ is the total number of
non-compact space-time dimensions.
We can diagonalize $M_p^2$ and absorb the diagonalizing 
matrices into the wave-function renormalization factor $Z^{1/2}(k)$ to express 
$M_p^2$ as a diagonal matrix.  These eigenvalues, which we shall denote by $m_{a,p}^2$,
are the squares of the physical masses. 
Taking into account the
non-diagonal nature of the wave-function renormalization factor $Z$, \refb{ek5pre} 
is modified to
\be \label{ek5}
S^{(n)}_{a_1\cdots a_n}(k_1,\cdots k_n) =
\lim_{k_i^2 \to -m_{a_i,p}^2} G^{(n)}_{b_1\cdots b_n}
(k_1,\cdots k_n)\prod_{i=1}^n \{Z_i^{-1/2}(k_i)_{a_i, b_i} (k_i^2 + m_{a_i,p}^2)\} \, ,
\ee
where $Z_i^{-1/2}$ is the inverse of the matrix $Z^{1/2}$ introduced in \refb{ek6} for the
$i$-th external state. In this expression we can interpret the sum over $b_i$'s as sum over
all fields in the theory if we define $Z^{1/2}(k)_{ab}$ and 
$Z^{-1/2}(k)_{ab}$
to be zero when $a,b$ label fields with different classical mass.

At tree level $Z=1$, $M_p^2=m^2 \, I$ and hence the $R^{(n)}$ defined in \refb{ek4} 
and $S^{(n)}$ defined in \refb{ek5} agree. 
In general however $R^{(n)}$ and $S^{(n)}$ are different.
While $S^{(n)}$ defined in \refb{ek5}
is the physically relevant quantity, string theory directly computes $R^{(n)}$ defined
in \eqref{ek4}.
Thus the question arises: how can we use string theory to compute on-shell S-matrix elements
beyond tree level? At a more basic level: how can we use string theory to calculate the
physical mass $m_{a_i,p}$ of the $i$-th particle?

When the external strings represent massless gauge particles, the situation improves
dramatically.
In this case gauge symmetry prevents mass renormalization and hence we have
$m_{a_i,p}^2=m_{a_i}^2=0$. As a result $R^{(n)}$ and $S^{(n)}$
differ only by the wave-function renormalization factor $Z$. This can be fixed by
using analyticity property of the S-matrix, {\it e.g.} the S-matrix should factorize into the
product of lower point S-matrices when the external momenta are such that some
internal line could become on-shell. Thus string world-sheet computation can be used to
compute the S-matrix of massless external states.

Now typically in string theories many massive string states 
appear as one particle intermediate states
in the scattering of massless states and as a result the S-matrix of the massless states
can have poles when the square of appropriate combination of external momenta 
approaches the squared mass of a massive state. The location of this pole gives
information about the mass of the massive state while the residue at this pole contains
information about the S-matrix involving massive external states. 
However this procedure does not always work. Some string theories contain massive
states which do not appear as one particle
intermediate states in the scattering of massless particles.
We shall now describe some examples of such situations.

\begin{enumerate}

\item Consider bosonic string theory compactifed on a circle $S^1$. In this case a state
carrying a winding number (and/or momentum) 
along $S^1$ cannot be produced as single particle intermediate state in the
scattering of massless states which do not carry any momentum and winding 
charge.\footnote{Such states could still appear in pairs in the intermediate channel,
producing a cut in the S-matrix of massless states, and by examining where the cut
begins, we can find the mass of the intermediate state.
But it is much harder to identify cuts
than poles in the S-matrix, and we shall not explore this option.}

\item Another
notable example is SO(32) heterotic string theory which contains massive states
belonging to the spinor representation of SO(32). They cannot appear as single
particle intermediate states in the scattering of massless external states which are all
in the adjoint or singlet representation of SO(32). Thus the S-matrix element involving these
particles cannot be computed by examining any massless S-matrix element near its poles.
\end{enumerate}

In order to deal with these cases we shall
try to develop a different strategy -- compute the mass renormalization directly. 
We shall focus on a special class of states -- which we
shall call special states -- for which the analysis 
simplifies. We shall conclude this
section by describing these special states and their relevance to the problems mentioned above.

Let us suppose that we are dealing with a string theory with $D+1$ non-compact dimensions,
with $SO(D,1)$ Lorentz invariance. Then while discussing the
mass renormalization of  a massive state we can go to the rest frame of the particle so that
the spatial component $\vec k$ of the momentum vanishes.  
In this frame we consider physical states described by vertex operators of the form
\be \label{eprimary}
c\, \bar c\, e^{\pm {\bf i} k_0 X^0}\, V
\ee
where $c,\bar c$ are ghost fields and
$V$ is a dimension $(h,h)$ primary made of the compact coordinates and the
oscillators of the non-compact spatial coordinates. The on-shell condition on $k_0$ is
\be \label{emassshell}
k^0 = m, \quad m^2 = 4(h-1)\, ,
\ee
in the $\alpha'=1$ units.
The operators $V$ will form a finite dimensional representation of the 
$SO(D)$ little group.
If the world-sheet theory has additional global symmetry group $G$ associated with the compact
directions then the operators $V$ will also belong to finite dimensional representation of
this symmetry group.

Now consider all operators of the form $e^{\pm {\bf i}k_0 X^0} \OO$
where $\OO$'s are dimension $(h-1, h-1)$ operators made of the ghost fields,
compact coordinates and oscillators of $X^0$ and 
the non-compact spatial coordinates. 
They can be organised into irreducible representations of $SO(D)\times G$.
Among them the operators which are not of the form \refb{eprimary} will be called 
unphysical vertex operators at mass level $m$.\footnote{Technically 
the unphysical operators described here can
the divided into two kinds, BRST trivial ones and states which are not invariant under BRST
transformation. The former are called pure gauge and the latter are called unphysical. We shall
not need to make this distinction, and call all such states unphysical.}
We shall define
{\it special vertex operators} to be a set of vertex operators of the form given in
\refb{eprimary} belonging to those irreducible representations of the symmetry group 
$SO(D)\times G$ such that there are no unphysical vertex operator 
at mass level $m$
transforming
in these representations. Put another way, if the unphysical vertex operators at mass level
$m$ transform
in certain irreducible representations $R_1,R_2,\cdots$ then the special vertex operators
are those physical states which transform in representations other than $R_1, R_2,\cdots$.
In this case the two point function of any special vertex operator and
an unphysical operator on any Riemann surface will vanish.

We shall now give some examples of special vertex operators.
\begin{enumerate}
\item Consider bosonic string theory in $25+1$ dimensions. We consider vertex operators of
the form \refb{eprimary} with $V$ given by
\be \label{especialv}
S\left[ \p X^{i_1} \cdots \p X^{i_n} \bar \p X^{j_1} \cdots \bar \p X^{j_n}\right]\, .
\ee
where $X^i$ for $1\le i\le 25$ are the spatial coordinates and $S$ denotes the operation of
taking the symmetric
traceless part of the product. This belongs to a rank $2n$ symmetric traceless representation of
$SO(25)$ -- also known as the leading Regge trajectory. In order to get
an unphyical state at the same mass level we have to replace some of the $\p X^i$ or $\bar \p X^j$
by ghost or $X^0$
excitations and/or replace the product of some of the $\p X^i$'s and/or 
$\bar\p X^i$'s by higher derivatives
of $X^i$'s. This clearly reduces the rank of the tensor and hence the unphysical states
cannot belong to the rank $2n$ symmetric tensor representation of $SO(25)$. Thus 
vertex operators of the form \refb{especialv} are special.
\item Consider bosonic string theory compactified on a circle. Let $Y$ be the coordinate along the
compact direction, and $Y_L$ and $Y_R$ be its left and right-moving components on the
world-sheet. We now consider the vertex operator of the form \refb{eprimary} with
\be \label{especial2}
V= e^{\pm {\bf i} (n/R - w R) Y_L/2} e^{\pm {\bf i} (n/R + w R) Y_R/2}\, 
S\left[ \bar \p X^{i_1} \cdots \bar \p X^{i_p} \p X^{j_1} \cdots \p X^{j_q}\right]\, ,
\quad p-q = nw\, ,
\ee
where $X^i$ for $i=1,\cdots 24$ denote the non-compact directions and $S$ stands for the
projection into rank $p+q$ symmetric traceless representation of $SO(24)$. Following the same
argument as before it follows that there are no unphysical vertex operators at this mass level
carrying $n$ units of momentum, $w$ units of winding and belonging to the rank $p+q$ 
symmetric
traceless representation of $SO(24)$. Thus these are also special states according to the definitions
given above.
\item Finally we note that the stable non-BPS states of SO(32) heterotic string theory --
which correspond to the lowest mass states in the spinor representation of SO(32) -- are also
special states. Besides the ghost fields and the $e^{\pm {\bf i} k_0 X^0}$ factor, the left-moving part
of the vertex operator is given by the SO(32) spin field of dimension 2, 
and has no further oscillator excitations.
Level matching requires that the right-moving part of the Neveu-Schwarz (NS) sector vertex operator
corresponds to level 3/2 excitations above the NS-sector ground state. We can take this to be
$\psi^i \psi^j \psi^k$ where $\psi^i$ for $1\le i\le 9$ are the world-sheet superpartners of the
9 non-compact bosonic coordinates. This belongs to the totally anti-symmetric rank 3 tensor
representation {\bf 84} of SO(9). It is easy to see that any other unphysical state at this mass
level, obtained by replacing some of the $\psi^i$'s by ghost or $\psi^0$ 
oscillators or derivatives of
$\psi^i$ or bosonic coordinates
cannot belong to the {\bf 84} representation of $SO(9)$. Thus these states are special states.
\end{enumerate}

The reader might have noticed that there is a close relationship between special states which are
prevented from mixing with the unphysical states due to global symmetry on the world-sheet
and the
states which cannot appear as poles in the scattering of massless states due to conserved charges.
Indeed the lowest mass states in 
each of the examples of the latter kind given earlier also correspond to special states.
On general grounds one expects that 
in every charge sector we can construct a set of special states by saturating the required
oscillator levels by (anti-)symmetric products of bosonic (fermionic) fields associated with the
non-compact coordinates. For this reason we shall focus on computation of physical mass and
S-matrix elements involving these special states and massless states, since the renormalized mass
and S-matrix elements of all other states can be obtained from the locations of the poles of the
S-matrix involving the special states and massless states.

\sectiono{Mass renormalization} \label{smass}

If we work in the rest frame, then the off-shell continuation of a special vertex operator
would correspond to deforming $k^0$ away from $m$. This keeps the vertex operator 
primary but it no longer has dimension 0. Thus in order to define the correlation functions
of such vertex operators on a Riemann surface we need to make a choice of local
coordinate around every point on the Riemann surface. If $z$ denotes some reference 
cordinate system on the
Riemann surface then the local coordinate $w$ around some point $z=z_0$ is described by some 
function $f(w;z_0)$ 
that maps  the $w$ plane to the $z$ plane around $z=z_0$.
We take $f(0; z_0)=z_0$ and $f(w;z_0)$ to be analytic around $w=0$.
Thus $f$ depends on both, the choice of the reference coordinate system $z$ and
the choice of the local coordinate system $w$.
The vertex operator at $z_0$ is inserted using the local coordinate $w$, which corresponds
to inserting its conformal transformation by the function $f(w;z_0)$ in the $z$ coordinate
system\cite{Nelson}. Thus if the off-shell vertex operator is a primary operator of 
dimension $(\delta, \delta)$
then we multiply it by $(f'(0;z_0))^\delta (\overline{f'(0;z_0)})^\delta$ while inserting it
into the
correlation function in the $z$ coordinate system, $f'$ being the derivative of the function 
$f(w;z_0)$
with respcet to $w$. For more general vertex operator representing general off-shell string state
the same procedure would work although the conformal transform of the vertex operator will be more
complicated.\footnote{For related approaches to defining off-shell amplitudes in string theory,
see \cite{Cohen:1985sm,Cohen:1986pv,AG1,AG2}}.
This definition makes the correlation function invariant under a change of
reference coordinate system $z$, but dependent on the choice of local coordinates $w$, \i.e.\ the
function $f(w;z_0)$. As a result, if we define off-shell strng amplitudes by integrating such
correlation functions over the moduli space, the result will depend on
the choice of local coordinates.

In a sense the situation in string theory is not very different from that in a gauge theory. In
gauge theory for computing the mass renormalizaton of a massive charged particle we have to
first compute the off-shell propagator carrying momentum $k=(k^0,\vec 0)$ 
and then look for its poles in the
$k^0$ plane. The off-shell propagator is not gauge invariant; however the location of its pole
in the $k^0$ plane is gauge invariant and leads to a gauge invariant definition of the
renormalized mass.
Thus a possible strategy in string theory will be to 
consider off-shell propagator that depends on the choice of local coordinates,  
look for its poles
in the $k^0$ plane and prove that the location of the pole is independent of the choice of 
local coordinates even
though the propagator itself is not gauge invariant. If we had an underlying string field 
theory then this analysis will be parallel to that in an ordinary gauge theory. 
This can be done in principle for bosonic string theory where a complete closed string field
theory is known\cite{9206084}.\footnote{Due to the presence of 
tachyon, the
mass renormalization in this case is infrared dvergent.} 
At present there is no known 
string
field theory for closed heterotic and superstring theories except a closed heterotic string field 
theory at tree level\cite{0409018}.
Nevertheless 
we can try to extract the
relevant features of the off-shell string theory amplitudes from a bosonic string field
theory and then develop a general proof of indpendence of the renormalized mass
of the choice of local coordinates that does not require the existence of an
underlying string field theory. The essential
features seem to be the following:
\begin{enumerate}
\item Bosonic string field theory gives a triangulation of the punctured Riemann surface 
equipped with local coordinate system at each puncture. Using this local coordinate system
we can define off-shell amplitudes.
\item Near boundaries of the moduli space where a Riemann surface of genus $n$ 
degenerates into two
Riemann surfaces of genus $n_1$ and $n_2=n-n_1$ connected by a long handle, the
choice of the local coordinates of the original Riemann surface matches with the choice
of the local coordinates of the lower genus surfaces. The precise meaning on `near
boundaries of the moduli space' will be made clear later (see item 6 in the discussion 
in \S\ref{sfinite} (above eq.\refb{edefbardelta}).
\end{enumerate}

For an off-shell amplitude induced from string field theory the above requirements are
automatically satisfied, but even in the absence of string field theory we could try to choose
local coordinates at the punctures consistent with the above criteria. Indeed even before the
construction of fully covariant closed string field theory, such choices of local coordinates were
explored (see {\it e.g.} \cite{divecchia}).
Given such a choice of local coordinates,
we can define off-shell two point functions in string theory and define the
mass to be the location of the pole in the $k^0$ plane. The important point is to show that
this definition is independent of the choice of local coordinates.

{}From now on we shall restrict our analysis to bosonic string theory. We discuss possible
generalizations to
superstring and heterotic string theories in \S\ref{sdiscuss}.

\subsection{Analysis of poles of off-shell two point function} \label{sanalysis}

Let us denote the set of all the special vertex operators by
$c\bar c V_i\, e^{{\bf i} k_0 X^0}$ and the corresponding states as 
\be
c_1 \bar c_1 |V_i\rangle \otimes |k^0, \vec k=0\rangle\, .
\ee
In the zero mode sector of non-compact bosons labelled by $(k^0,\vec k)$, the states 
satisfy the usual $\delta$-function normalization.
The operaors $V_i$ will be chosen so that
in the rest of the matter-ghost CFT,  they satisfy the
orthonormality relation
\be
\langle V_i| c_{-1} \bar c_{-1} c_0\bar c_0 c_1 \bar c_1 |V_j\rangle =\delta_{ij} \, .
\ee
Let $F(k)$ be the off-shell two point function of special states 
obtained by summing over all genera. If there are $n_p$ special states at mass level $m$ then
$F(k)$ is an $n_p\times n_p$ matrix satifying
\be
F(k) = F(-k)^T\, ,
\ee
where $F^T$ denotes transpose of $F$.
Then the off-shell propagator of special states 
is given by\footnote{We have removed an overall factor of $-{\bf i}$ and also absorbed a factor
of $-{\bf i}$ into the definition of $F(k)$.}
\be \label{efullprop}
{1\over k^2 + m^2} + \left({1\over k^2+m^2}\right)^2 \, F(k)\, ,
\ee
where $m$ is the tree level mass. The first term represents the tree level propagator whereas
the first factor of the second term is the effect of the two external propagators. 
$F(k)$ admits a genus expansion of the form $\sum_n F_n g^{2n}$ in string
coupling $g$, with higher genus contributions having higher order poles at $k^2+m^2=0$
from regions of the
moduli space where the Riemann surface degenerates into two or more
Riemann surfaces of lower genera connected by long handles, with the two external
vertices lying on the two  lower genus Riemann surfaces at the two ends.
We expect that after resummation, \refb{efullprop} may be written as
$Z^{1/2}(k)\, (k^2 + M_p^2)^{-1} (Z^{1/2}(-k))^T$ for some physical mass$^2$
matrix $M_p^2$ and wave-function renormalization matrix $Z^{1/2}(k)$ which has
no pole near $k^2=-m^2$. 
This will be seen explicitly in \refb{epole}-\refb{epole4} below.
We can take $M_p^2$ to be diagonal by absorbing the diagonalizing matrix into the definition
of $Z^{1/2}(k)$. If  $m_{a,p}^2$ for $a=1,2\cdots n_p$ 
are the eigenvalues of the mass$^2$ matrix $M_p^2$ then
the physical poles of the propagator are at $k^2=-m_{a,p}^2$.

Now consider the effect of changing the local coordinate system by an infinitesimal
amount. Let the change in $F$ to first order be $\delta F$. Then in order that the location
of the poles of the propagator in the $k^2$ plane 
does not shift, the net change in \refb{efullprop} must be of the form of an overall
multiplicative factor that renormalizes $Z^{1/2}(k)$.
Thus we require
\ben \label{erequire}
&&  {1\over k^2 + m^2} +  \left({1\over k^2+m^2}\right)^2 \, (F(k) +\delta\, F(k)) \nonumber \\
&=& (1 + \delta Y(k)) \left\{ {1\over  k^2 + m^2} + \left({1\over k^2+m^2}\right)^2 \, F(k)\right\}
(1 + \delta Y(-k))^T
\, ,
\een
for some matrix $\delta Y(k)$ whose genus expansion is free from any poles at
$k^2+m^2=0$. 
Equivalently we can write
\be \label{ereqd}
\delta F(k) = (k^2 + m^2) \, \delta Y(k) + (k^2 + m^2) \, (\delta Y(-k))^T + \delta Y(k) \, F(k)
+ F(k) \, \delta Y(-k)^T\, .
\ee

At each genus the two point function $\delta F$ receives two contributions
--  from the change of local coordinates at the vertex carrying momentum $k$ and the change
in local coordinates at the vertex carrying momentum $-k$. Both these contributions have an
explicit factor of $k^2+m^2$ due to the fact that when $k^2+m^2=0$ the vertex is on-shell
and hence there is no dependence on local coordinates. 
In concrete terms, since the off-shell vertex 
operator of a special state is a primary of dimension $((k^2+m^2)/4, (k^2+m^2)/4)$, if we insert
such an operator 
at the origin $w=0$ of the local cordinate system, and then
change the local coordinate from $w$ to $w+\epsilon(w)$ then we pick up a net multiplicative
factors of $(1+\epsilon'(0))^{(k^2 + m^2)/2} \simeq (1 + (k^2 + m^2) \epsilon'(0)/2)$.
Thus we introduce the function
$\delta H$ via the relations
\be
\delta F(k) =  (k^2+m^2) \, \delta H(k) + (k^2+m^2) \, (\delta H(-k))^T\, ,
\ee
where the first term is the effect of the change of local coordinates at the vertex carrying
momentum $k$ and the second term is the effect of change of local coordinates at the vertex
carrying momentum $-k$. The rules for computing $\delta H$ are the same as that of
$F$ except that at one of the punctures the vertex $c\bar c V_i$ is replaced by
$\epsilon'(0) c\bar c V_i /2$. 
We shall call the puncture where the effect of change of local coordiantes is inserted the
`special puncture'.
Eq.\refb{ereqd} can now be satisfied by choosing $\delta Y(k)$ such that  
\be \label{ewrite}
\delta H(k) = \delta Y(k) + (k^2 + m^2)^{-1} \, 
\delta Y(k) \, F(k)\, .
\ee
Our goal will be to show the existence of $\delta Y(k)$ satisfying
\refb{ewrite} such that the genus expansion of $\delta Y(k)$ does not have any pole at
$k^2+m^2=0$.

We now claim that there exist quantities $\wt F$ and $\delta \wt H$ with the properties that the
genus expansion of
neither of them has any poles near $k^2+m^2=0$, both have genus expansion starting
at one loop and $F$ and $\delta H$ can be expressed in terms of $\wt F$ and $\delta\wt H$
as
\be \label{eqF}
F = \wt F (1 - (k^2+m^2)^{-1} \wt F)^{-1}\, ,
\ee
\be \label{eqY}
\delta H = \delta \wt H (1 - (k^2+m^2)^{-1} \wt F)^{-1}\, . 
\ee
Let us first proceed assuming this to be true. 
From eqs.\refb{eqF} and \refb{efullprop} we see that the full propagator is given by
\be \label{epole}
(k^2+m^2)^{-1} + (k^2 + m^2)^{-2} \wt F \, (1 - (k^2+m^2)^{-1}\wt F)^{-1} 
= (k^2 + m^2 - \wt F(k))^{-1}\, .
\ee
If we choose a real basis of fields in position space then we have $\wt F(k)^\dagger=\wt F(k)$ and
$\wt F(k)^T = \wt F(-k)$. In this case 
by choosing suitable unitary matrix $U(k)$ satisfying $U(-k)^T=U(k)^\dagger$
we can express $\wt F(k)$ as 
$U(k) \wt F_d(k) U(k)^\dagger$ where $\wt F_d(k)$ is a diagonal matrix
satisfying $\wt F_d(-k)=\wt F_d(k)$. 
Furthermore the genus expansion of $U(k)$ is free from poles at $k^2+m^2=0$ since 
$\wt F(k)$ has this property.
We can now express \refb{epole} as
\be \label{epole1}
U(k) (k^2 + m^2 - \wt F_d(k))^{-1} U(k)^\dagger = U(k) (k^2 + m^2 - \wt F_d(k))^{-1} U(-k)^T\, . 
\ee
Let $M_p^2$ denote the diagonal matrix that describes the locations of the zeroes of the
eigenvalues of the diagonal
matrix $k^2 + m^2-\wt F_d(k)$ in the $-k^2$ plane.
We can solve for this iteratively starting with the leading order
solution $k^2=-m^2$. 
Then we can write  
\be \label{epole2}
(k^2 + m^2 - \wt F_d(k))^{-1} = X_d(k) (k^2 + M_p^2)^{-1}\, ,
\ee
where $X_d(k)$ is a diagonal matrix whose genus expansion 
does not have any pole near $k^2=-m^2$
and satisfies $X_d(-k)=X_d(k)$. Defining
\be \label{epole3}
Z^{1/2}(k) = U(k) \sqrt{X_d(k)}\, ,
\ee
satisfying $Z^{1/2}(k)^\dagger = Z^{1/2}(-k)^T$
we can express the propagator \refb{epole} as
\be \label{epole4}
Z^{1/2}(k) \, (k^2 + M_p^2)^{-1} \, Z^{1/2}(-k)^T\, .
\ee
The genus expansion of $Z^{1/2}(k)$ does not have any poles at $k^2+m^2=0$ since
neither $U(k)$ nor $X_d^{1/2}(k)$ has such poles.

Now
using  eq.\eqref{eqF}  we can 
express  \refb{eqY} as
\be
\delta H =  \delta \wt H   (1 + (k^2+m^2)^{-1} F) \, .  
\ee
Comparing this with \refb{ewrite}
we get
\be \label{efin}
\delta Y = \delta \wt H\, .
\ee
Since $\delta \wt H$ does not have any pole near $k^2+m^2=0$ this establishes that $\delta Y$
also does not have any pole near $k^2+m^2=0$. This in turn establishes the desired result that
the locations of the poles of \refb{efullprop} in the $k^2$ plane do not change under change in local
coordinates.

\subsection{Explicit construction of $\wt F$ and $\delta \wt H$} \label{sfinite}

It now remains to prove the existence of pole free $\wt F$ and $\delta \wt H$ satisfying 
\eqref{eqF} and \eqref{eqY}. 
We shall do this in steps.
\begin{enumerate}
\item 
First we extend the definitions of $\delta  H$ and $F$
where we allow the external states inserted at the punctures (except at the special puncture) 
to be general string states of ghost number two,\footnote{As will 
become clear later, we need to extend this
definition only to those states which are annihilated by $L_0-\bar L_0$, $b_0$ and $\bar b_0$.} 
inserted using the same
local coordinate system as before. This makes 
$F$ into an infinite dimensional square matrix which we shall call $\FF$ and $\delta  H$ into an
$n_p\times$ infinite dimensional matrix (since one of its two punctures is special) 
which we shall call $\delta  \HH$. 

\item
We now use another insight from
string field theory\cite{9206084}: it provides us with a triangulation of the moduli space in which 
the full moduli space of a genus $n$ Riemann surface with two punctures
can be decomposed into a `one particle
irreducible' region $R_n$ and the rest. The region $R_n$ has the property that
it does not contain any
boundary of the moduli space in which a genus $n$ Riemann surface degenerates into a
pair of two punctured lower genus Riemann surfaces connected by a long handle, with each side
containing one of the original punctures. The rest of the moduli space is 
obtained by gluing in all possible ways lower genus puctured
Riemann surfaces  corredponding to
regions $R_{n'}$ by the
plumbing fixture procedure\cite{divecchia,Polchinski:1988jq}.
If we denote by $\wh\FF$ and $\delta  \wh \HH$ the contributions to $\FF$ and 
$\delta \HH$ from
integration over the  one particle
irreducible regions $R_n$ of the moduli spaces, then $\wh \FF$ and
$\delta  \wh\HH$ have no poles at $k^2+m^2=0$ since the region of integration does not
include the degenerating Riemann surfaces.  We shall shortly discuss how to define 
$\wh \FF$ and
$\delta  \wh\HH$ in the absence of a string field theory underlying the choice of local
coordinates we have made.
There is also an additional subtle point in the definition of $\delta\wh \HH$ which will be discussed
in point \ref{eeight} of this discussion.

\item We can regard $\wh\FF$  and $\FF$ as maps from $\HH\times \HH$ to ${\mathbb C}$ where $\HH$
denotes the space of string states of ghost number 2. However
since string states of ghost number 4 form the dual vector space of string states of ghost
number 2 via the inner product
in the CFT, we can also regard $\FF$ and $\wh\FF$ as maps from states of
ghost number 2 to string states of ghost number 4. 
We shall in fact include left multiplication by the operator $\bar b_0  b_0$ -- the zero modes of
the $b$ and $\bar b$ ghost fields -- to regard $\FF$ and
$\wh\FF$ as maps from states of ghost number 2 to states of ghost number 2.
This is the way we shall interpret $\FF$ and
$\wh\FF$ from now on. By including similar factor in the definition of
$\delta\HH$ and $\delta\wh \HH$, they 
can be regarded as maps from string states of ghost number
2 to the space of special states. 

\item With this convention the full contribution to $\FF$ and $\delta  \HH$
is obtained by gluing $\wh \FF$ and
$\delta  \wh\HH$ using the string propagator
\be \label{eDelta}
\Delta  = {1\over 4\pi} \int_0^{2\pi} d\theta 
\, \int_0^\infty ds\, e^{-s (L_0+\bar L_0) +i\theta(L_0-\bar L_0)}=
 {1\over 2}\, \delta_{L_0, \bar L_0}\, \int_0^\infty ds\, e^{-s (L_0+\bar L_0)}\, .
\ee
The normalization of $\Delta$ has been chosen such that acting on special states at
mass level $m$ it gives $(k^2+m^2)^{-1}$.
We can now express $\FF$ and $\delta \HH$ as
\ben  \label{effhh}
\FF = \wh \FF + \wh \FF \Delta \wh \FF + \wh \FF \Delta \wh \FF \Delta \wh \FF+\cdots
= \wh\FF (1-\Delta\wh\FF)^{-1} = (1 - \wh \FF\Delta)^{-1} \wh\FF\, , \nonumber \\
\delta  \HH = \delta  \wh\HH + \delta  \wh\HH \Delta \wh \FF + \delta  \wh\HH \Delta \wh \FF \Delta \wh \FF
+\cdots
= \delta\wh\HH (1-\Delta\wh\FF)^{-1}\, .
\een
Note that each factor of $\Delta$ is accompanied by a hidden factor of $\bar b_0 b_0$
coming from $\wh\FF$; these are required to provide the correct integration measure on the
moduli space. Eqs.\refb{effhh} provide us with explicit implementation of plumbing fixture,
building a higher genus Riemann surface from gluing of lower genus punctured Riemann
surfaces. 

In the world-sheet description, $\wh\FF \Delta \wh\FF$ contains integration over those Riemann
surfaces, which can be obtained by gluing two Riemann surfaces corresponding to
regions of the moduli space included in the definition of $\wh\FF$, at one each
of their punctures by the relation
\be \label{egluing}
w_1 w_2 = e^{-s+ i\theta}, \quad 0\le s<\infty, \quad 0\le \theta < 2\pi\, ,
\ee
where $w_1$ and $w_2$ are the local coordinates at the punctures. Similar interpretation
holds for terms like $\delta\wh\HH \Delta \wh\FF$.

\item In the absence of an underlying string field theory we can use \refb{effhh} to define
$\wh\FF$ and $ \delta  \wh\HH$. Consider for example $\wh\FF$. 
Let $\FF_n$ and $\wh\FF_n$ denote the genus $n$ contribution to $\FF$ and
$\wh\FF$ respectively.
Since both $\FF$ and
$\wh \FF$ have genus expansion beginning at genus one, the genus expansion
of \refb{effhh} tells us that
$\wh\FF_1$ is identical to
$\FF_1$. Now at genus two the right hand of the first equation in \refb{effhh} gets a
contribution from the $\wh\FF_1\Delta \wh\FF_1$ term. This represents integration over 
certain region of the genus two moduli space with the same integrand as that in the expression
for $\FF_2$. Then $\wh\FF_2$ is given by the integral of the same integrand over the
complementary region of the genus two moduli space. The same process can now be repeated
for higher genus,  $\wh\FF_n$ being given by an integration over certain region
of the genus $n$ moduli space with the same integrand as that of $\FF_n$. The region of
integration is the region that is not covered by gluing the lower genus $\wh\FF_m$'s by
$\Delta$. By construction $\wh \FF_n$ defined this way does not include integration over any
region of the moduli space that 
corresponds to degeneration of the Riemann surface of the kind discussed before, since
these regions are already included from the gluing of lower genus contributions.
Since the structure of the second equation in \refb{effhh}
is similar to that of the first equation, the genus $n$ contribution to $\delta  \wh\HH$
will be given by integration over the same region of the genus $n$ moduli space as
that for $\wh\FF_n$, with the integrand being the same as that of $\delta  \HH$.

\item \label{psix}
Note however that for this procedure to be consistent it is essential that for those Riemann 
surfaces which are built by gluing lower genus Riemann surfaces, represented in the genus
expansion of the right hand side of 
\refb{effhh} by product of lower genus contributions connected by $\Delta$, the choice of
local coordinates at the punctures must coincide with those on the lower genus Riemann
surfaces. We shall assume that the local coordinates have been chosen this way even if they
are not inherited from an underlying string field theory. We also need to assume that the
Riemann surfaces produced by the gluing procedure are all distinct, \i.e.\ the same Riemann
surface should not be produced by two different gluing procedure. This can be achieved with
an appropriate choice of local coordinates, {\it e.g.} by scaling the local coordinates by a 
sufficiently small number $\lambda$ we can ensure that the gluing produces only Riemann
surfaces close to degeneration and hence different gluing produces different Riemann surfaces.

\item
We define $P_T$ to be the projection operator into all states of momentum $k$ -- physical and
unphysical -- with  $L_0=\bar L_0 = (k^2 + m^2)/4$,
and define
\be \label{edefbardelta}
\bar\Delta \equiv \Delta - {1\over k^2+m^2} P_T\, ,
\ee
\ben \label{edefffdbar}
\bar\FF \equiv \wh \FF + \wh \FF \bar\Delta \wh \FF + \wh \FF \bar\Delta \wh \FF \bar\Delta \wh \FF+\cdots
= \wh\FF (1-\bar\Delta \wh\FF)^{-1} = (1 - \wh\FF \bar\Delta)^{-1} \wh\FF\, , \nonumber \\
\delta  \bar\HH \equiv \delta  \wh\HH 
+ \delta  \wh\HH \bar\Delta \wh \FF + \delta  \wt\HH\bar \Delta \wh \FF \bar\Delta \wh \FF+\cdots
=\delta\wh\HH(1-\bar\Delta \wh\FF)^{-1}\, .
\een
Physically $\bar\FF$ and $\delta\bar\HH$ 
denote `one particle irreducible' contribution to appropriate two point functions
of fields at mass level $m$ {\it after integrating out the fields at other mass levels}. 
Using \refb{edefffdbar} we can rewrite
\eqref{effhh} as
\ben \label{efirst}
\FF &=& \bar\FF (1- (k^2+m^2)^{-1} P_T \bar\FF)^{-1} \nonumber \\
&=&
\bar \FF + \bar \FF (k^2+m^2)^{-1} P_T \bar \FF + \bar \FF (k^2+m^2)^{-1} P_T \bar \FF (k^2+m^2)^{-1} P_T \bar \FF+\cdots
\, , \nonumber \\
\delta  \HH &=& \delta\bar\HH (1- (k^2+m^2)^{-1} P_T \bar\FF)^{-1} \nonumber \\
&=& \delta  \bar\HH + \delta  \bar\HH (k^2+m^2)^{-1} P_T \bar \FF + \delta  \bar\HH (k^2+m^2)^{-1} P_T 
\bar \FF (k^2+m^2)^{-1} P_T \bar \FF+\cdots 
\, . \nonumber \\
\een

\item
We now define
\be \label{eprojection2}
P = c_1 \bar c_1 |V_i\rangle\langle V_i| c_{-1} \bar c_{-1} c_0\bar c_0 \otimes I_{zero}
\, ,
\ee
as the projection operator into the special states with tree level mass $m$. 
Here $I_{zero}$ corresponds to
identity operator acting on the zero mode sector of non-compact bosons, labelled by
$(k^0,\vec k)$. 
In the following we shall omit explicit mention of the operator $I_{zero}$ as the 
various operators we shall work with will always act as identity operator in this sector.
Applying the projection operator $P$ on both sides of the first equation 
in \refb{efirst} and from the right
in the second equation in \refb{efirst}, and noting that
\be
P \,\FF \, P = F, \quad \delta \HH \, P = \delta  H, 
\ee
we get
\ben \label{efirstsecond}
F = P\bar \FF\, P + P \, \bar \FF (k^2+m^2)^{-1} P_T \bar \FF \, P+ P\, \bar \FF (k^2+m^2)^{-1} P_T \bar \FF (k^2+m^2)^{-1} P_T \bar \FF\, P+\cdots
\, , \nonumber \\
\delta  H = \delta  \bar\HH \, P+ \delta  \bar\HH (k^2+m^2)^{-1} P_T \bar \FF\, P + \delta  \bar\HH (k^2+m^2)^{-1} P_T 
\bar \FF (k^2+m^2)^{-1} P_T \bar \FF \, P+\cdots
\, . \nonumber \\
\een
\item Now $P$ denotes projection operator into special states which transform in certain representations
of the symmetry group $SO(D)\times G$. $P_T-P$ denotes projection operator into states at the
same mass level which are not special, and hence by definition transform in representations 
of $SO(D)\times G$ other
than those in which special states transform. Thus the two point function of special and non-special
states on any Riemann surface vanishes, leading to $(P_T-P)\wh\FF P=0$,
$(P_T-P)\Delta P=0$. This in turn gives
\be \label{eptp}
 (P_T-P)\, \bar \FF \, P=0, \qquad P \, \bar\FF \, (P_T-P)=0\, .
\ee
Using this we can replace the $P_T\bar \FF \, P$ factors in \refb{efirstsecond} by
$P\, \bar \FF \, P$. Defining 
\be \label{edefagain}
\wt F = P \, \bar \FF \, P, \quad \delta  \wt H = \delta  \bar \HH \, P,
\ee
we get
\ben
F &=& \wt F + \wt F (k^2+m^2)^{-1} \wt F + \wt F (k^2+m^2)^{-1} \wt F (k^2+m^2)^{-1} \wt F + \cdots
\nonumber \\
\delta  H &=& \delta  \wt H + \delta  \wt H(k^2+m^2)^{-1} \wt F+ \delta  \wt H(k^2+m^2)^{-1} \wt F
(k^2+m^2)^{-1} \wt F + \cdots \, .
\een
\item
This reproduces \eqref{eqF}, \eqref{eqY}.
Furthermore since  $\bar\FF$ and $\delta  \bar\HH$ have no poles at $k^2+m^2=0$ it follows that 
$\wt F$ and $\delta  \wt H$ defined in \refb{edefagain}
also have no poles at $k^2+m^2=0$. This is the desired result.

\item  \label{eeight}
We now come to a subtle point in the definition of $\delta\wh \HH$ 
alluded to earlier. First consider the contributions $\delta\wh \HH$ to the right hand
side of the second equation in \refb{effhh}.
Naively, $(k^2+m^2)\delta\wh \HH$ 
represents the difference between two contributions, both given by integrating over
the same subspace of the moduli space that is
used to define $\wh\FF$. In one of them 
we use the original
local coordinate encoded in the function $f$ at the puncture carrying momentum $k$,
while in the other one we use the local coordinates encoded in the
function $f+\delta f$ at the puncture carrying momentum $k$. 
This difference is clearly what we need to compute the contribution to 
$(k^2+m^2)\delta\HH$ from these
Riemann surfaces. For reasons that will become clear soon, let us denote this contribution
to $\delta\wh\HH$ by $\delta_0\wh\HH$.

Now consider the contribution  $(k^2+m^2)
\delta_0\wh\HH \Delta \wh\FF$. Again this gives the difference
between two contributions: $B-A$. The first contribution $A$ is
obtained by gluing the Riemann surfaces corresponding to $\wh\FF$ to those corresponding to
$\wh\FF$ at one each of their punctures using the original coordinate system $f$, with the coordinate
at the external punctures also given by the original local coordinate system $f$.
This induces a specific local coordinate system at the external punctures  
on the Riemann surfaces represented by $\wh \FF \Delta\wh\FF$ (see {\it e.g.}
\cite{peskin}).
By the compatibiity condition discussed in point \ref{psix}, this 
is the correct choice of coordinate system on the Riemann surfaces  
in the original system.
The second contribution $B$ 
is 
obtained by gluing Riemann surfaces represented by $\wh\FF$ and
$\wh\FF$ at one each of their punctures using the original coordinate system $f$, with the coordinate
at the external puncture carrying momentum $k$ given by the deformed local coordinate system 
$f+\delta f$.
This induces a specific local coordinate system at the external punctures carrying momentum
$k$ on the Riemann surfaces represented by $\wh \FF \Delta\wh\FF$, but this is not the
correct choice of coordinate system as prescribed in the deformed system since we are still
using the original local coodinate system $f$ for the gluing. 
Let $f+\delta_1 f$ denote the coordinate at the external puncture carrying
momentum $k$ that we get using the 
gluing procedure described above
and $f+\delta  f$ be the local coordinate at the external puncture for the deformed
system which we would get by using the coordinate system $f+\delta f$ both for
external puncture and for the punctures we are using for gluing.\footnote{Two points
may need clarification here. First we are using the same symbol $f$ for the coordinates 
on the component Riemann surface and the Riemann surface we get by gluing these
components since $f$ stands for the original
choice of local coordinates on {\it all} Riemann surfaces. Similar remark applies to $f+\delta f$.
The second point is that while comparing the coordinate systems $f+\delta f$ and 
$f+\delta_1 f$ we work at the same point in the moduli space of the glued Riemann surface.}
Let us denote by $(k^2+m^2)
\delta_1\wh\HH$ the difference between
the two contributions, the $(k^2+m^2)$ factor being there due to the fact that the external vertex
represents a dimension (0,0) primary in the $k^2+m^2\to 0$ limit, and hence $f+\delta_1 f$ and
$f+\delta f$ acting on the external vertex gives the same result in the $k^2+m^2\to 0$ limit.
Then
$(k^2+m^2)\delta_0\wh\HH \Delta \wh\FF+(k^2+m^2)\delta_1\wh\HH$ gives the desired 
difference between the off-shell
amplitudes computed using the deformed system and the original system. 
We can then
add the error term $\delta_1\wh\HH$ to $\delta_0\wh\HH$
to define a corrected $\delta\wh \HH$ so that the net contribution to $\delta\HH$ can still
be written as the right hand side of the second equation in \refb{effhh}.  
The only possible caveat with this is that since the definition of
$\delta_1\wh\HH$ involves integration over moduli 
spaces of Riemann surfaces corresponding to $\wh\FF\Delta\wh\FF$, this
involve a degeneration limit
where the parameter $s$ appearing in the definition of $\Delta$ in \refb{eDelta}
goes to $\infty$. 
Integration over $s$ from this region could produce a pole at $k^2=-m^2$.
We shall now argue that this does not happen. 
For this note that in the $s\to\infty$ limit the Riemann surface degenerates into two
Riemann surfaces, and the local coordinates induced at external punctures
are inherited from the local coordinates at the external punctures of the Riemann surfaces
which are being glued, and independent of the local coordinates at the punctures which we
use to glue the two Riemann surfaces. Thus the functions
$f+\delta f$ and $f+\delta_1 f$ should  become identical as $s\to\infty$.
From this we conclude that for large $s$ they should
differ by a term proportional to $q=e^{-s+i\theta}$. 
As a result the expression for
$\delta_1\wh\HH$, which involves difference in the contributions with local coordinates
$f+\delta_1 f$ and $f+\delta f$ at the external puncture carrying momentum $k$, will have
an extra factor of $q$ and/or $\bar q$ in the integrand.
Since the leading contribution to the
integrand in \refb{eDelta} 
in the $s\to\infty$ limit comes from states of mass level $m$ and is proportional to
$e^{-(k^2+m^2) s/2}\sim |q|^{(k^2+m^2)/2}$, 
we see that an extra factor of $q$ and/or $\bar q$ 
in the
integrand will kill the pole at $k^2=-m^2$. Thus $\delta_1\wh \HH$ is free from poles
at $k^2+m^2=0$.

To summarize, 
$(k^2+m^2)(\delta_1\wh \HH+ \delta_0\wh \HH \Delta\wh\FF)$,
added to $ P\wh\FF\Delta\wh\FF$, produces correctly the contribution to off-shell
Green's function with the deformed coordinate system from those Riemann surfaces which
correspond to $\wh\FF\Delta\wh\FF$. 
Furthermore $\delta_1\wh\HH$ does not contain any pole at $k^2=-m^2$. Defining
$\delta\wh\HH=\delta_0\wh\HH+\delta_1\wh\HH$, we ensure the equality of two sides of
the second equation of \refb{effhh} to this order. 
We can then move on to the term
$(\delta_1\wh\HH+\delta_0\wh \HH \Delta\wh\FF)\Delta\wh\FF$ and carry out  similar 
analysis, generating
further correction $\delta_2\wh\HH$ to $\delta\wh\HH$.
After carrying out this procedure to the desired order in perturbation theory we can ensure
that  \refb{effhh} and hence all subsequent equations still hold with this new definition of
$\delta\wh\HH$.

\end{enumerate}

\sectiono{S-matrix elements} \label{smatrix}

The on-shell S-matrix element for massive external string states can be analyzed by
following a procedure similar to the one used for mass renormalization. 
Again we shall restrict to S-matrix elements of special states (and possibly massless states
for which there is no mass renormalization); the S-matrix elements of other states
can be found in principle from the above by computing its residues at appropriate poles.
Using the
given local coordinate system for $n$-punctured Riemann surfaces we compute the
off-shell $n$-point function $F^{(n)}_{a_1\cdots a_n}(k_1, \cdots k_n)$ of $n$ external legs. 
With the help
of \refb{ek5}, \refb{ek4} we can then define the on-shell S-mtatrix elements via
\be \label{esmatrix}
S^{(n)}_{a_1\cdots a_n}(k_1,\cdots k_n) =
\lim_{k_i^2 \to -m_{a_i,p}^2} F^{(n)}_{b_1\cdots b_n}
(k_1,\cdots k_n)\prod_{i=1}^n \left\{Z_i^{-1/2}(k_i)_{a_ib_i} (k_i^2 + m_{a_i,p}^2)
\, (k_i^2+m_{a_i}^2)^{-1}\right\}  \, .
\ee
We shall now prove that $S^{(n)}$ defined this way is invariant under a change of 
local coordinates even though $F^{(n)}$'s themselves transform under such changes.
The change in $S^{(n)}$ comes from two sources: the change in $F^{(n)}$ and the change in
$Z_i^{-1/2}(k_i)$. We begin by computing the change in $Z_i^{-1/2}(k_i)$. 
First of all comparing \refb{ek6} with the transformation law \refb{erequire} of the
propagator under a change of local coordinates,
we get
\be \label{erhs}
\delta Z_i^{1/2}(k_i) = \delta  Y_i(k_i)  Z_i^{1/2} (k_i) \, ,
\ee
where $\delta  Y_i$ is the same as $\delta Y$ 
introduced in \refb{erequire} and computed in \refb{efin}
for the $i$-th external state.
The multiplication on the right hand side of \refb{erhs}
should be regarded as a matrix multiplication.
This gives
\be \label{ezhalf}
\delta Z_i^{-1/2}(k_i) = - Z_i^{-1/2}(k_i) \delta  Y_i(k_i)  = - Z_i^{-1/2}(k_i) \delta  \wt H_i(k_i) \, ,
\ee
where in the last step we have used the equality of $\delta Y$ and $\delta \wt H(k)$
given in  \refb{efin}.

Next we shall analyze the contribution to $\delta F^{(n)}_{b_1\cdots b_n}$. 
This can be expressed as
\be
\delta F^{(n)}_{b_1\cdots b_n} = \sum_j \delta_j F^{(n)}_{b_1\cdots b_n}\, ,
\ee
where $\delta_j$ denotes the effect of the change of local coordinates at the $j$-th
puncture. 
We shall later show that there exist quantities
$ \wt F_{j;b_1\cdots b_n}^{(n)}$ and $\delta_j \wt H^{(n)}_{b_1\cdots b_n}$
whose perturbation expansions have no poles at $k_j^2+m_{a_j}^2=0$ and in terms
of which we have the relations
\be \label{eidep}
F^{(n)}_{b_1\cdots b_n} = \left( 1 - (k_j^2+m_{a_j}^2)^{-1} \wt F_j(k_j)\right)_{b_jc_j}^{-1} 
\wt F^{(n)}_{j,b_1\cdots b_{j-1}c_j b_{j+1}\cdots b_n}\, ,
\ee
and
\ben \label{eidelfn}
\delta_j F^{(n)}_{b_1\cdots b_j} &=& (k_j^2 +m_{a_j}^2) \left[ \delta_j \wt H^{(n)}_{b_1\cdots b_n} 
+ \delta  \wt H_j(k_j)_{b_jc_j}
\left( 1 - (k_j^2+m_{a_j}^2)^{-1} \wt F_j(k_j)\right)^{-1}_{c_jd_j} \right. \nonumber \\ &&
\qquad \qquad \qquad \qquad  \left .(k_j^2+m_{a_j}^2)^{-1} 
\wt F^{(n)}_{j,b_1\cdots b_{j-1} d_j b_{j+1}\cdots b_n} \right]\, ,
\een
where the quantities $\wt F_j(k_j)$ and $\delta  \wt H_j(k_j)$
are the same matrices which were called $\wt F(k_j)$ and $\delta  \wt H(k_j)$ in
eqs.\refb{eqF}, \refb{eqY}, with the subscript $j$ indicating that we have to use appropriate 
matrices $(\wt F_j(k_j))_{b_jc_j}$ and $(\delta  \wt H_j(k_j))_{b_jc_j}$ relevant for
the $j$-th external leg. The various products and inverses appearing in
\refb{eidep}, \refb{eidelfn} are then interpreted as matrix products and matrix
inverses acting on the $j$-th leg.

We shall prove the existence of $ \wt F_{j;b_1\cdots b_n}^{(n)}$ 
and $\delta_j \wt H^{(n)}_{b_1\cdots b_n}$ with the desired properties later;
for now we shall proceed assuming this to be true. 
Using \refb{eidep} we can express \refb{eidelfn} as
\be \label{edeltafjntwo}
\delta_j F^{(n)}_{b_1\cdots b_j} 
=(k_j^2 +m_{a_j}^2)  \, \delta_j \wt H^{(n)}_{b_1\cdots b_n} + \delta  \wt H_j(k_j)_{b_jc_j} 
\, F^{(n)}_{b_1\cdots b_{j-1} c_j b_{j+1}\cdots b_n}\, .
\ee
We are now in a position to calculate $\delta S^{(n)}$. Using eqs.\refb{esmatrix}, 
\refb{ezhalf} and
\refb{edeltafjntwo} we get
\ben
\delta S^{(n)}_{a_1\cdots a_n} 
&=& \lim_{k_i^2 \to -m_{a_i,p}^2 \, \forall \, i}  \sum_{j=1}^n  \prod_{\ell=1\atop  \ell\ne j}^n \left\{Z_\ell^{-1/2}(k_\ell)_{a_\ell b_\ell} (k_\ell^2 + m_{a_\ell,p}^2)
\, (k_\ell^2+m_{a_\ell}^2)^{-1}\right\} (k_j^2 + m_{a_j,p}^2)
\, (k_j^2+m_{a_j}^2)^{-1} \nonumber \\ &&
\qquad \qquad \left[\delta Z_j^{-1/2}(k_j)_{a_jb_j} F^{(n)}_{b_1\cdots b_j} +  Z_j^{-1/2}(k_j)_{a_jb_j} 
\delta_j F^{(n)}_{b_1\cdots b_n}
\right] \nonumber \\
&=& \lim_{k_i^2 \to -m_{a_i,p}^2 \, \forall \, i}  \sum_{j=1}^n  \prod_{\ell=1}^n 
\left\{Z_\ell^{-1/2}(k_\ell)_{a_\ell b_\ell} (k_\ell^2 + m_{a_\ell,p}^2)
\, (k_\ell^2+m_{a_\ell}^2)^{-1}\right\} \nonumber \\ &&
\times \sum_{j=1}^n  
\left[ -\delta  \wt H_j(k_j)_{b_jc_j}F^{(n)}_{b_1\cdots b_{j-1}c_jb_{j+1}\cdots b_n} 
+ (k_j^2 +m_{a_j}^2)  \delta_j \wt H^{(n)}_{b_1\cdots b_n} \right.
\nonumber \\ &&
\qquad \qquad 
\left. + \delta  \wt H_j(k_j)_{b_jc_j}F^{(n)}_{b_1\cdots b_{j-1}c_jb_{j+1}\cdots b_n}
\right]\nonumber \\
&=& \lim_{k_i^2 \to -m_{a_i,p}^2 \, \forall \, i} \, 
\prod_{\ell=1}^n \left\{Z_\ell^{-1/2}(k_\ell)_{a_\ell b_\ell} (k_\ell^2 + m_{a_\ell,p}^2)
\right\} 
\sum_{j=1}^n  \prod_{\ell=1\atop \ell\ne j}^n (k_\ell^2 +m_{a_\ell}^2)^{-1}\,
 \delta_j \wt H^{(n)}_{b_1\cdots b_\ell} \, 
 \, .
\een
Now note that the genus expansion of the
$j$-th term in the sum has no poles at $k_j^2+m_{a_j}^2=0$ since there
is no explicit factor of $(k_j^2+m_{a_j}^2)^{-1}$ and the genus expansion of $\delta_j \wt H^{(n)}$ 
does not contain any poles at $(k_j^2+m_{a_j}^2)=0$. 
As a result after resummation this term will have no pole at $k_j^2+m_{a_j,p}^2=0$, and after
being multiplied by the $(k_j^2 + m_{a_j,p}^2)$ term,  will
give vanishing contribution in the $k_j^2 \to -m_{a_j,p}^2$ limit. Since this analysis can be 
repeated for every $j$, we see that $\delta S^{(n)}$ vanishes. Thus the S-matrix is invariant under
a change in the local coordinates.

It remains to prove the existence of $ \wt F_{j;b_1\cdots b_n}^{(n)}$ 
and $\delta_j \wt H^{(n)}_{b_1\cdots b_n}$ satisfying \refb{eidep}, \refb{eidelfn} and having no poles
at $k_j^2=-m_{a_j}^2$ in their genus expansion. 
For this we first define $\FF^{(n)}_j$ by  allowing the $j$-th external state 
of $F^{(n)}$ to be an arbitrary string state. 
We also use the fact that the change in local coordinates generates a vertex proportional to
$(k^2+m^2)$ to introduce the quantity $\delta_j H^{(n)}$ via
\be\label{efnhn}
\delta_j H^{(n)}_{b_1\cdots b_n} = (k_j^2+m_{a_j}^2)^{-1} \, \delta_j F^{(n)}_{b_1\cdots b_n}\, .
\ee
Then in the same spirit as the $\wh\FF$ and $\delta\wh\HH$ defined in \refb{effhh} 
we introduce $\wh\FF^{(n)}_j$ and $\delta_j \wh H^{(n)}$ via the expansion:
\ben \label{erelations}
\FF^{(n)}_j&=&\wh\FF^{(n)}_j+ \wh \FF \Delta \wh\FF^{(n)}_j + \wh \FF \Delta \wh \FF \Delta
\wh\FF^{(n)}_j +\cdots = (1-\wh\FF\Delta)^{-1} \, \wh \FF^{(n)}_j\, ,\nonumber \\
\delta_j H^{(n)} &=& \delta_j \wh H^{(n)} + \delta\wh\HH_j \Delta \wh\FF^{(n)}_j 
+ \delta\wh\HH_j \Delta \wh \FF \Delta \wh\FF^{(n)}_j
+  \delta\wh\HH_j \Delta \wh \FF \Delta \wh \FF \Delta \wh\FF^{(n)}_j + \cdots \nonumber \\
&=& \delta_j \wh H^{(n)} +\delta\wh\HH_j \Delta 
(1-\wh\FF\Delta)^{-1} \, \wh \FF^{(n)}_j\, ,
\een
where $\wh\FF$ has been defined via \refb{effhh} and
$\delta\wh\HH_j$ is the same as $\delta\wh\HH$ defined in \refb{effhh}, but for
the $j$-th external state.
All multiplications in \refb{erelations} are matrix multiplications on the $j$-th external leg with
fixed indices $b_i$ for $i\ne j$ on all other legs. 
$\wh\FF^{(n)}_j$ and $\delta_j \wh H^{(n)}$ represent contributions to
$\FF^{(n)}_j$ and $\delta_j  H^{(n)}$ which are one particle irreducible on the
$j$-th external leg. Thus they are given by integration over 
subregions of the moduli space
of Riemann surface with the same integrands as $\FF^{(n)}$ and $\delta_j H^{(n)}$,
and these subregions have the property that they do not include any degeneration of the
$j$-th external leg.\footnote{The definition of $\delta_j \wh H^{(n)}$ 
suffers from subtleties of the
same kind that affects the definition of $\delta \wh \HH$, and these are dealt with in the same
way as in the case of $\delta\wh\HH$, following the
procedure discussed in  point \ref{eeight} at the end of \S\ref{smass}.}
Thus the genus expansions of 
$\wh\FF^{(n)}_j$ and $\delta_j \wh H^{(n)}$ do not have any pole at $k_j^2+m_{a_j}^2=0$. Now we
define
\ben \label{erelations2}
\bar\FF^{(n)}_j&=&\wh\FF^{(n)}_j+ \wh \FF \bar\Delta \wh\FF^{(n)}_j + \wh \FF \bar\Delta \wh \FF 
\bar\Delta
\wh\FF^{(n)}_j +\cdots = (1-\wh\FF\bar\Delta)^{-1} \, \wh \FF^{(n)}_j
\nonumber \\
\delta_j \wt H^{(n)} &=& \delta_j \wh H^{(n)} + \delta\wh\HH_j \bar\Delta \wh\FF^{(n)}_j 
+ \delta\wh\HH_j \bar\Delta \wh \FF \bar\Delta \wh\FF^{(n)}_j
+  \delta\wh\HH_j \bar\Delta \wh \FF \bar\Delta \wh \FF \bar\Delta \wh\FF^{(n)}_j + \cdots
\nonumber \\ 
&=& \delta_j\wh H^{(n)} + \delta \wh\HH_j \bar\Delta (1-\wh\FF\bar\Delta)^{-1} \wh\FF^{(n)}_j 
= \delta_j\wh H^{(n)} + \delta \wh\HH_j (1-\bar\Delta\wh\FF)^{-1}\bar \Delta  \wh\FF^{(n)}_j  \, ,
\een
where $\bar\Delta$ has been defined in \refb{edefbardelta}. 
Since $\bar\Delta$ has no poles at $k_j^2+m_{a_j}^2=0$, the genus expansions of
$\bar\FF^{(n)}_j$ and
$\delta_j \wt H^{(n)}$ also do not have any poles at $k_j^2+m_{a_j}^2=0$.
Using \refb{edefffdbar}, \refb{erelations} and \refb{erelations2} we get
\ben \label{etwoeq}
\FF^{(n)}_j&=&
(1 - \bar\FF (k_j^2+m_{a_j}^2)^{-1}P_T)^{-1} \bar \FF^{(n)}_j 
 \nonumber \\
\delta_j H^{(n)} &=& \delta_j \wt H^{(n)} + \delta \bar \HH_j (k_j^2+m_{a_j}^2)^{-1} P_T
(1 - \bar \FF (k_j^2+m_{a_j}^2)^{-1} P_T)^{-1} \bar \FF^{(n)}_j \, .
\een
We now define 
\be \label{enewdef}
\wt F^{(n)}_j = P \bar\FF^{(n)}_j\, .
\ee
Since the genus expansion of 
$ \bar\FF^{(n)}_j$  has no poles at $k_j^2=-m_{a_j}^2$, the genus expansion of
$\wt F^{(n)}_j$ also has no poles at $k_j^2=-m_{a_j}^2$.
It follows from the definition of special states that $\delta\bar\HH_j P_T 
= \delta\bar\HH_j P$. Using this and \refb{enewdef}, 
multiplying the first equation of \refb{etwoeq} by $P$ from the
left, using $P\FF^{(n)}_j= F^{(n)}$ and eqs.\refb{eptp},
\refb{edefagain} we can write the two equations in
\refb{etwoeq} as
\ben \label{etwoeqnew}
F^{(n)}&=&(1- (k_j^2+m_{a_j}^2)^{-1} \wt F_j)^{-1} \, \wt F^{(n)}_j 
 \, , 
 \nonumber \\
\delta_j H^{(n)} &=& \delta_j \wt H^{(n)} + \delta\wt H_j (k_j^2+m_{a_j}^2)^{-1} 
(1- (k_j^2+m_{a_j}^2)^{-1} \wt F_j)^{-1} \wt F^{(n)}_j \, .
\een
This reproduces \refb{eidep} and \refb{eidelfn} after using \refb{efnhn}.

\sectiono{Discussion and generalizations} \label{sdiscuss}

In this paper we have given an algorithm for computing renormalized mass and S-matrix
elements for a special class of massive states in bosonic string theory, and have shown that
these are independent of the specific off-shell continuation that we use for computing them.
While the results are in the same spirit as the proof of gauge invariance of physical mass
and S-matrix elements in a gauge theory, in many sense the analysis here is simpler than
in gauge theories. In the latter the gauge invariance results from cancellation between the
contributions from different Feynman diagrams, while here we do not require any such
cancellations. In fact if we had been trying to prove gauge invariance of renormalized mass
and S-matrix elements in string field theory, we would still need cancellation between 
different Feynman diagrams.\footnote{A change of local coordinates correspond to a field 
redefinition of the string field\cite{9301097} followed by a gauge transformation that is needed to bring the
transformed fields to the Siegel gauge.}

The simplicity in string theory of course is a consequence of the fact that in string theory there
is only one contribution from every genus. Technically the difference between our analysis and the
corresponding analysis in string field theory can be traced to the fact that in string field theory
a change in local coordinates will change the local coordinates not only at the external 
punctures, but also at the internal punctures that we use to glue two Riemann surfaces using
the plumbing fixture procedure. As a result each Feynman diagram gets additional contribution from the
change in local coordinates at the internal punctures which cancel between different Feynman 
diagrams. 

Clearly there are many generalizations of our analysis that are needed. 
We expect that for special vertex operators our 
analysis can be generalized in a straightforward manner to heterotic and superstring
theories. Consider bosonic states coming from the NS sector (in heterotic
string theory) or NS-NS sector (in type II string theories) -- for Ramond sector we expect
similar analysis to go through with the propagator \refb{eDelta} acquiring extra numerator
factors\cite{1209.5461}. 
In this case the choice of local coordinates at the punctures will have to be replaced by a
choice of local superconformal coordinates. On-shell vertex operators are independent of local
superconformal coordinates, and hence under a change of local coordinates 
off-shell  vertex operators change by a term proportional to $(k^2+m^2)$ as in the
case of bosonic string theory. 
We also have the analog of the gluing relations \refb{egluing} (see {\it e.g.}
\cite{1209.5461}) and hence the relations
\refb{effhh} with the bosonic propagator $\Delta$ 
given by the same expression as \refb{eDelta}.\footnote{By choosing the local 
coordinate system appropriately
we can ensure that 
the splitting of the moduli space we have used, {\it e.g.} in defining $\wh \FF$ etc,
requires only information about the region of the moduli space near the boundary where
complications arising out of non-splitness of the supermoduli space are not 
present\cite{1304.7798}.
For example by scaling the function $f(w;z_0)$ introduced at the beginning of
\S\ref{smass} to $f(\lambda w; z_0)$ for sufficiently small $\lambda$,
we can ensure that $\wh\FF$ includes the contribution from most of the
region of the moduli space except those close to the boundaries where the Riemann surface
splits apart into two lower genus surfaces.}  Thus we expect that extending our analysis to
superstring and heterotic string theories is straightforward.

For general external states we expect new
complications even in the bosonic string theory. 
This is due to the fact that under quantum correction the physical states would
begin mixing with the unphysical states and we need to take into account this mixing for defining
an appropriate off-shell continuation. 
For example from genus two onwards $\wh\FF$ will have non-zero matrix element
between a physical state and a BRST trivial state from the boundary of the region of integration
of the moduli space that defines $\wh\FF$, forcing us to change the definition of the physical state.
Furthermore the required mixing will
 depend on the particular off-shell continuation we choose \i.e.\ on the choice of local
 coordinates at the punctures. We expect that once these effects are taken into account, 
 we shall be able to directly prove that the renormalized mass and S-matrix elements are
 independent of the off-shell continuation for all physical states, suitably defined. 

In fact it seems to us that the off-shell formalism could be a useful way of studying 
string perturbation
theory both for massive and massless external states, and can be used to give alternate proofs of
well known results in string theory. For example in the standard on-shell approach the proof of
decoupling of pure gauge states, corresponding to trivial elements of the BRST cohomology,
involves first showing that the result is given by a total derivative in the moduli space and then
showing that the boundary terms arising from the integration of the total derivative terms vanish.
In the off-shell formalism the boundary terms can be ignored altogether since they can be made to
vanish by appropriate off-shell continuation of the external momenta. The price we pay
is that due to BRST non-invariance of the external off-shell states there will be additional terms 
proportional to one or more powers of $(k_i^2+m_{a_i}^2)$ associated with the external states.
In individual terms these may be cancelled by inverse powers of $(k_i^2+m_{a_i}^2)$ coming from
integration over moduli near the boundaries. Thus the task 
will be to show that the final result vanishes nevertheless in the on-shell limit.

\bigskip

{\bf Acknowledgement:}
We thank Rajesh Gopakumar, Michael Green and Barton Zwiebach for useful discussions.
The work of R.P. and A.S. was
supported in part by the 
DAE project 12-R\&D-HRI-5.02-0303. 
A.R. was supported by the Ramanujan studentship of Trinity College, Cambridge and would like to thank HRI, Allahabad for hospitality during the initial stages of this work.
The work of A.S. was also supported in
part by the
J. C. Bose fellowship of 
the Department of Science and Technology, India.

\end{document}